\documentclass{article}

\usepackage{arxiv}
\usepackage{authblk}
\usepackage[utf8]{inputenc} 
\usepackage[T1]{fontenc}    
\usepackage{hyperref}       
\usepackage{url}            
\usepackage{booktabs}       
\usepackage{amsfonts}       
\usepackage{nicefrac}       
\usepackage{microtype}      
\usepackage{lipsum}
\usepackage{graphicx}
\usepackage{amsmath}
\usepackage{amssymb}
\usepackage[table,xcdraw]{xcolor}
\usepackage{multirow}
\usepackage{framed,multirow}

\usepackage{amssymb}
\usepackage{latexsym}
\usepackage{amsmath}
\usepackage{url}
\usepackage{xcolor}
\usepackage{float}
\usepackage{color}
\usepackage[normalem]{ulem}

\usepackage{amssymb}
\usepackage{pifont}
\newcommand{\cmark}{\ding{51}}%
\newcommand{\xmark}{\ding{55}}%

\usepackage{hyperref}
\title{Longitudinal detection of new MS lesions \\using Deep Learning} 

\author[1]{Reda Abdellah Kamraoui}
\author[1]{Boris Mansencal}
\author[3]{José V Manjon}
\author[1]{Pierrick Coupé}

\affil[1]{Univ. Bordeaux, Bordeaux INP, CNRS, LaBRI,\\ UMR5800, PICTURA, F-33400 Talence, France}
\affil[2]{Univ. Bordeaux, INSERM, Neurocentre Magendie,\\
 U1215, F-3300 Bordeaux, France\\}
\affil[3]{ITACA, Universitat Politècnica de València,\\
 46022 Valencia, Spain\\}

\begin{document}
\maketitle

\begin{abstract}
The detection of new multiple sclerosis (MS) lesions is an important marker of the evolution of the disease. The applicability of learning-based methods could automate this task efficiently. However, the lack of annotated longitudinal data with new-appearing lesions is a limiting factor for the training of robust and generalizing models.
In this work, we describe a deep-learning-based pipeline addressing the challenging task of detecting and segmenting new MS lesions.
First, we propose to use transfer-learning from a model trained on a segmentation task using single time-points. Therefore, we exploit knowledge from an easier task and for which more annotated datasets are available.
Second, we propose a data synthesis strategy to generate realistic longitudinal time-points with new lesions using single time-point scans. 
In this way, we pretrain our detection model on large synthetic annotated datasets.
Finally, we use a data-augmentation technique designed to simulate data diversity in MRI. By doing that, we increase the size of the available small annotated longitudinal datasets.
Our ablation study showed that each contribution lead to an enhancement of the segmentation accuracy. Using the proposed pipeline, we obtained the best score for the segmentation and the detection of new MS lesions in the MSSEG2 MICCAI challenge. 

\end{abstract}

\section{Introduction}
Multiple sclerosis (MS) is a chronic autoimmune disease of the central nervous system.
The pathology is characterized by inflammatory demyelination and axonal injury, which can lead to irreversible neurodegeneration.
The disease activity, such as MS lesions, can be observed using magnetic resonance imaging (MRI).
The detection of new MS lesions is one of the important biomarkers that allow clinicians to adapt the patient's treatment and assess the evolution of this disease. 

Recently, the automation of single time-point MS lesion segmentation has shown encouraging results. Many techniques showed performance comparable to clinicians in controlled evaluation conditions (see \cite{carass2017longitudinal,commowick2016msseg}). These methods use a single time-point scan to segment all appearing lesions at the time of the image acquisition.
However, these cross-sectional techniques are not adapted to the longitudinal detection of new lesions. Indeed, using these methods requires repeatedly running the segmentation process for each time-point independently to segment MS lesions before detecting new ones. Unlike the human reader, these methods are not designed to jointly exploit the information contained at each time point.
Consequently, single-time MS lesion segmentation methods performance is not optimal for the detection of new lesions between two time-points.
Moreover, inconsistencies may appear between segmentations of both time-points since they are processed independently.

To specifically address this detection task using both time-points at the same time, some detection methods have been proposed. 
In one of the earliest works, \cite{bosc2003automatic} used a nonlinear intensity normalization method and statistical hypothesis test methods for change detection. \cite{elliott2013temporally} used a bayesian tissue classifier on the time-points to estimate lesion candidates followed by a random-forest-based classification to refine the identification of new lesions.  \cite{ganiler2014subtraction} used image subtraction and automated thresholding. \cite{cheng2018multi} integrated neighborhood texture in a machine learning framework. \cite{salem2018supervised} trained a logistic regression model with features from the image intensities, the image subtraction values, and the deformation field operators. \cite{schmidt2019automated} used lesion maps of different time-points and FLAIR intensities distribution within normal-appearing white matter to estimate lesion changes. \cite{kruger2020fully} used a 3D convolutional neural network (CNN) where each time-point is passed through the same encoder. Then, the produced feature maps are concatenated and fed into the decoder.

Training learning-based methods for the task of new lesions detection require a dataset specifically designed for the task. The most obvious form of the training data would be a longitudinal dataset of MS patients (with two or more successive time-points) with new appearing lesions carefully delineated by experts in the field. 
However, the construction of such a dataset is very difficult.
To begin, new lesions may take several months or even years to appear and be visible in a patient's MR image.
Moreover, a time-consuming and costly process is necessary for several experts to annotate new lesions from the two time-points and to obtain an accurate consensus segmentation.
Although the organizers of the MICCAI Longitudinal Multiple Sclerosis Lesion Segmentation Challenge (MSSEG2-challenge \cite{challenge_dataset}) provided such a dataset, the training set is severely impacted by class imbalance (see Section \ref{dataset_imbalance} for more details) due to the difficulty of finding new lesions in the follow-up scan.
This under-representation of new lesions in longitudinal datasets is limiting the training of state-of-the-art deep learning algorithms from scratch on this complex task. Besides, achieving generalizing results on unseen domains (see \cite{maartensson2020reliability,bron2021cross,omoumi2021buy}) may requires more data diversity.

Several studies tackled the problem of training data scarcity. 
First, transfer learning is a strategy used to create high-performance learners trained with more widely available data from different domains when the target domain/task data are expensive or difficult to collect (see \cite{torrey2010transfer, weiss2016survey}).
Second, synthetic data generation is performed by using a model able to simulate realistic artificial data that can be used during training (see \cite{tremblay2018training, Tripathi_2019_CVPR, khan2021brain}).
Third, data-augmentation is a set of techniques used to handle the variability in real-world data by enhancing the size and quality of the training dataset (see \cite{shorten2019survey}). Recently, \cite{zhang2020generalizing} showed that applying extensive data augmentation during training also enhances the generalization capability of the methods.

In this paper, we propose an innovative strategy integrating these three strategies into a single pipeline for new MS lesion segmentation to tackle data rarity for our task.
First, we use transfer-learning to exploit the larger and more diverse datasets available for the task of single-point MS lesion segmentation which does not require longitudinal data.
Second, we propose a novel data synthesis technique able to generate two realistic time-points with new MS lesions from a single FLAIR scan.
Third, we use a data-augmentation technique to simulate a large variety of artifacts that may occur during the MRI acquisitions.

\section{Method and Material}

\begin{figure*}[t]
    \centering
    \newcommand{\sz}{1\textwidth}
    \includegraphics[width=\sz]{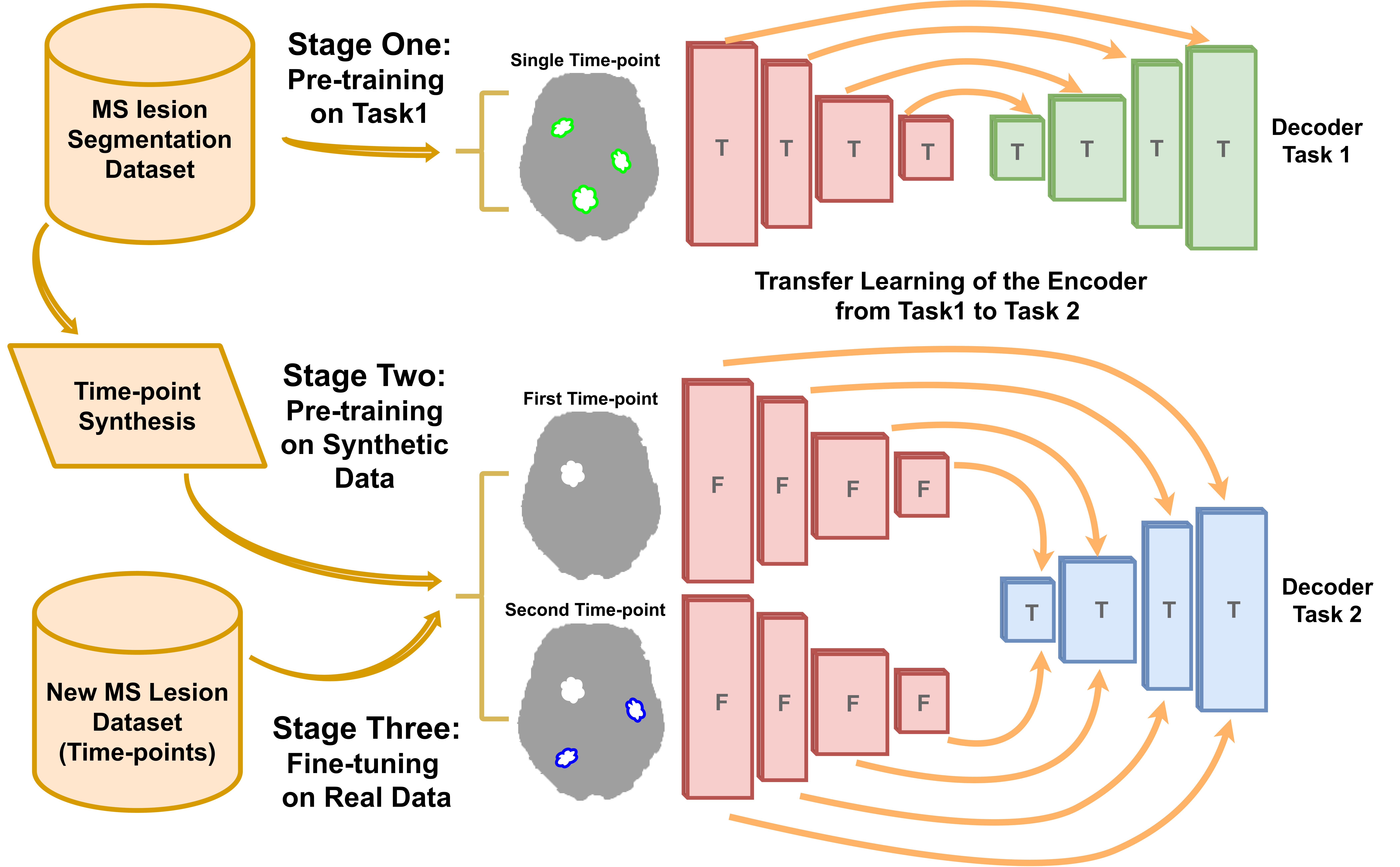}
      \caption{The pipeline of our new MS lesion segmentation method. The three stages include: First, the pre-training on the task of single-time-poing MS lesion segmentation (Task 1). Second, pre-training on the task of new MS lesions segmentation (Task 2) with synthetic data. Third, fine-tuning the model with real data. The encoder weights are trained (T) in Stage One and freezed (F) in Stage Two and Stage Three.  }
    \label{fig:abstract_pipeline}
\end{figure*}

\begin{figure*}[t]
    \centering
    \newcommand{\sz}{1\textwidth}
    \includegraphics[width=\sz]{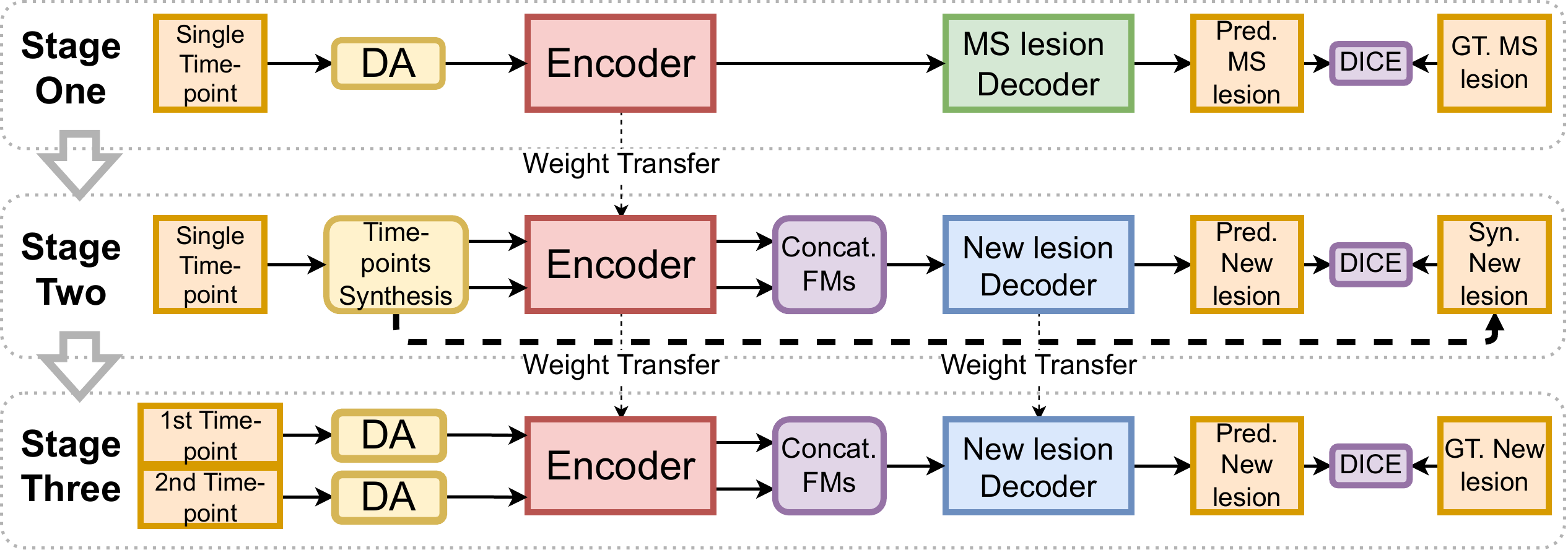}
      \caption{The diagram represents our training method. Input images are augmented with the proposed method (DA). The encoder trained in Stage One is used in Stage Two and Stage Three to extract feature maps (FMs) of the two-time points. The aggregation block (Concat. FMs) is used to combine features }
    \label{fig:pipeline}
\end{figure*}

\subsection{Method Overview}
To deal with data rarity for new MS lesion segmentation, we proposed a three stage pipeline as shown in Fig. \ref{fig:abstract_pipeline}. 
\textbf{In Stage One}, an encoder-decoder network is trained on the task of single time-point MS lesions segmentation. This step aims to train the encoder part of the network to extract relevant features related to MS lesions that can be used in the next steps. Stage One enables to indirectly use large datasets dedicated to single time-point MS lesion segmentation for the task of new lesions segmentation. This stage is detailed in Section \ref{encoder_pretraining}.
\textbf{In Stage Two}, the new lesions segmentation model composed of the previous task encoder is pretrained with synthetic data. To this end, we trained external models able to generate two realistic time-points from a single image also taken from single time-point MS datasets. It combines the effects of lesion inpainting and lesion generating models to simulate the appearance of new lesions. This strategy is detailed in Section \ref{synthesis}.
\textbf{In Stage Three}, the decoder is fine-tuned with real longitudinal data from the new MS lesion training-set of the MSSEG2 MICCAI challenge. 

\subsection{Transfer-learning from single time-point MS lesion segmentation task}\label{encoder_pretraining}

The encoder used for new MS lesion segmentation is first trained on single time-point lesion segmentation (see Fig.\ref{fig:pipeline}, from Stage One to Stage Two). This choice is motivated by two reasons. First, we consider that datasets for MS lesion segmentation with lesion mask segmentation by experts are more diverse and larger than available datasets for new lesions segmentation (which requires a longitudinal study). Second, the task of MS lesion segmentation is tightly close to the one of new MS lesion segmentation. By learning to segment lesion, the model implicitly learns the concept of a lesion, either the lesion is considered new or was already existing in the first time-point.   
To conclude, since there is a proximity between the two tasks, there is likely a gain from exploiting the large amount of training data of the first task to improve the second task's performance. 

\subsubsection{Model Architecture Design}\label{design}
\begin{figure}[htb!]
  \centering
  \newcommand{\sz}{\textwidth}
  \includegraphics[width=\sz]{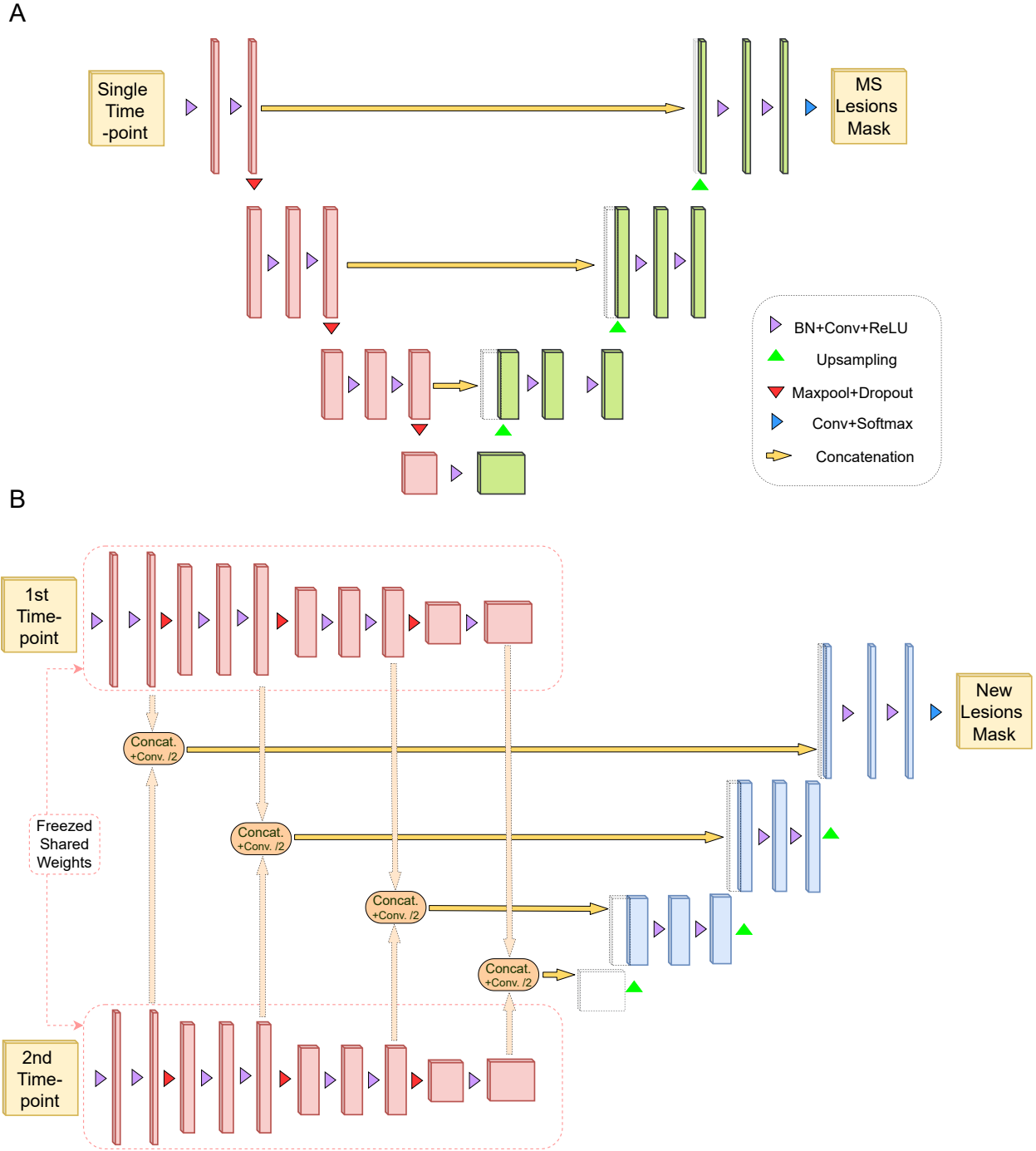}
  \caption{Part A represents U-Net like architecture composed of an encoder (in red) and a decoder for the task of MS lesion segmentation (in green). This task requires a single time-point as input and produce the MS lesion mask. Part B shows a siamese-encoder (in red) to extract the same sets of features from the two time-points. Same-level features are aggregated with a combination module and are forwarded to a decoder for the task of new lesions segmentation (in blue).  }
  \label{fig:archi}
\end{figure}

Our method is based on the transfer learning from the task of “Single time-point MS lesion segmentation” to the task of “new lesions segmentation from two time-points”. Thus two different architectures are used but with the same building blocks for each task. 
For the first task, a 3D U-Net shape architecture is used, as shown in Fig. \ref{fig:archi} Part A. This kind of architecture has been very effective and robust for MS lesion segmentation \cite{kamraoui2022deeplesionbrain, isensee2021nnu}.
It is composed of an encoder and a decoder linked with one another by skip connections.

For the second task, a siamese-encoder followed by a single decoder is used, as shown in Fig.\ref{fig:archi} Part B. The shared-weights encoders are chosen to extract the same set of features from both time points. Then, these features resulting from the different levels of both encoder paths are aggregated (see Fig. \ref{fig:archi} Part B). The aggregation module is composed of concatenation and a convolution operation. Feature maps are first concatenated by channels (\textit{i.e.} result channel size is twice the original size), then the convolution operation aggregates the information back to the original channel size. Finally, the aggregated features are passed through the decoder.

\subsection{Time-points Synthesis}\label{synthesis}

\begin{figure*}[t]
    \centering
    \newcommand{\sz}{1\textwidth}
    \includegraphics[width=\sz]{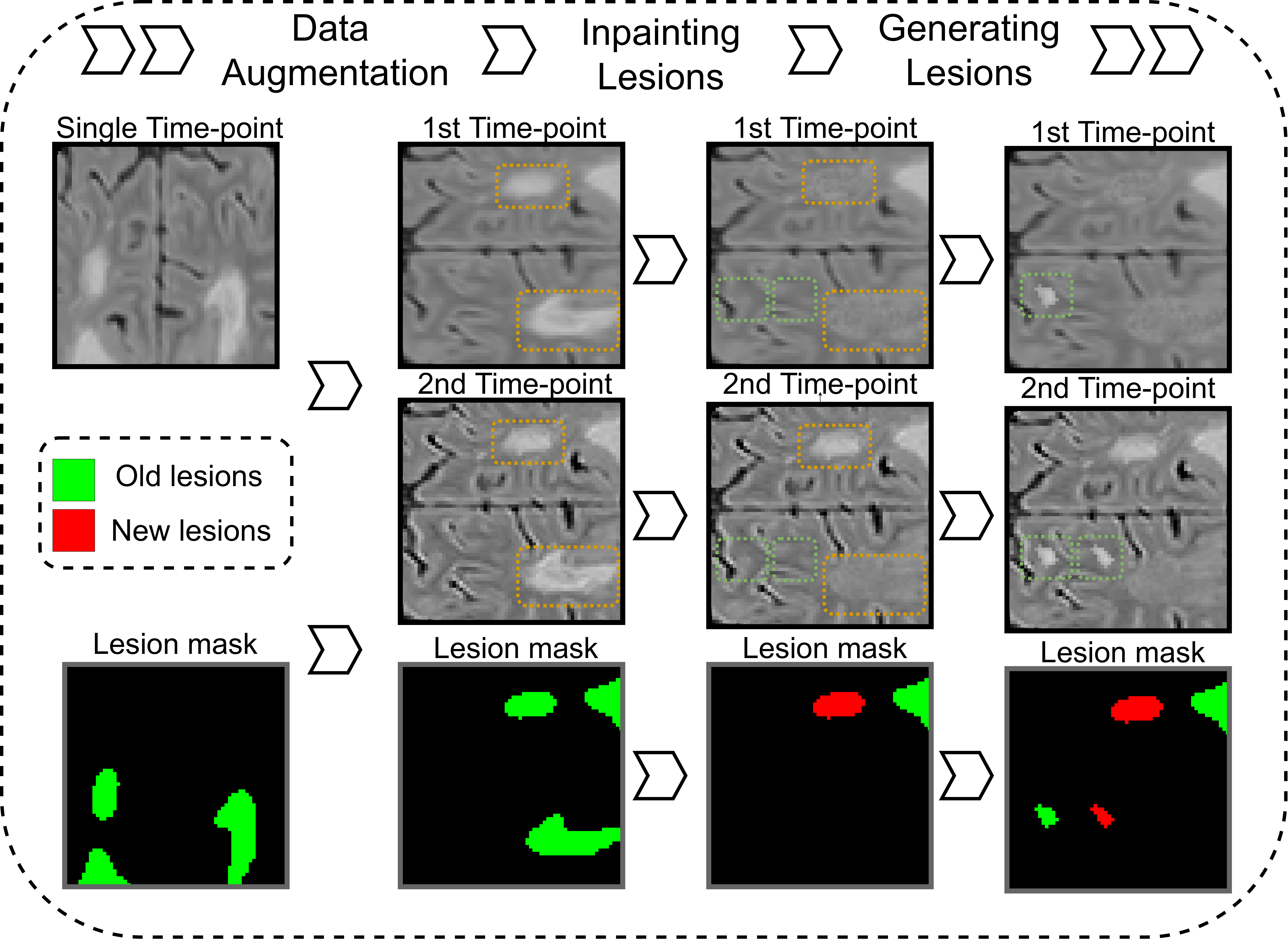}
      \caption{Synthetic time points with new MS lesion generation pipeline. Dashed orange and green rectangles on images represent areas where lesion are inpainted or generated }
    \label{fig:synthesis}
\end{figure*}

The data synthesis method is based on the simulation of new MS lesions between two time-points using single time-point FLAIR images. As shown in Fig. \ref{fig:synthesis}, our pipeline generates “on the fly” synthetic 3D patches that represent longitudinal scans of the same patient with an evolution in their lesion mask.
The synthetic data is generated in three steps. 
In the first step, a 3D FLAIR patch and its MS lesion segmentation mask are randomly sampled from different MS lesion segmentation datasets (see Section \ref{single_dataset}). Then, the patch and lesion mask are randomly augmented with flipping and rotations. A copy of the FLAIR patch is performed to represent the two time-points. Then, both identical patches are altered with the described data augmentation (see Section \ref{IQDA}) to differentiate the two patches. At this point, the lesion masks of the two synthetic time-points are still identical. Thus, there are no new lesions.  
In the second step, a connected component operation is used to separate each independent lesion from the lesion mask. Each lesion is either inpainted (\textit{i.e.} removed) from one of the two time-points or both of them, or it can be kept in both of the time-points. The lesion inpainting model is used to inpaint the lesion region with hallucinated healthy tissue (see Section \ref{Lesion Inpainting Model}). Next, the new lesion mask is constructed from lesion regions that have been kept in the second time-point but not the first one.
In the third step, the lesion generator model is used to simulate new synthetic lesions at realistic locations (using white/gray matter segmentation and a probabilistic distribution of MS lesions on the brain in the MNI space). Synthetic lesions are generated for one of the time-points or both of them (see Section \ref{Lesion Generator Model}). Similar to the previous step, the new lesion mask is updated to include only the generated lesions on the second time-point.

\subsubsection{Lesion Inpainting Model}\label{Lesion Inpainting Model}
The lesion inpainting model is trained, independently and priorly to our proposed pipeline, with randomly selected 3D FLAIR patches which do not contain MS lesions or white matter hyperintensities. 
Similarly to \cite{manjon2020blind}, 
A 3D U-Net network is optimized to reconstruct altered input images. Specifically, the input patch is corrupted with Gaussian noise (\textit{i.e.} with a mean and a standard deviation of the image intensities) in lesion-like areas at random locations. 
When the model is trained, it can be used to synthesize healthy regions in lesion locations that are replaced with random gaussian (see \cite{manjon2020blind} for details).   

\subsubsection{Lesion Generator Model} \label{Lesion Generator Model}
The lesion generator is trained before our proposed pipeline to simulate realistic lesions.
The generator is a 3D U-Net network with two input channels and one output channel. The first input channel receives an augmented version of 3D FLAIR patches containing MS lesions where lesions are replaced with random noise. The second input channel receives the MS lesion mask of the original 3D FLAIR patch. The output channels predict the original 3D FLAIR patch with lesions. Thus, the trained model can simulate synthetic MS lesions from a 3D patch of FLAIR and its corresponding lesion mask.

\begin{figure*}[t]
    \centering
    \newcommand{\sz}{1\textwidth}
    \includegraphics[width=\sz]{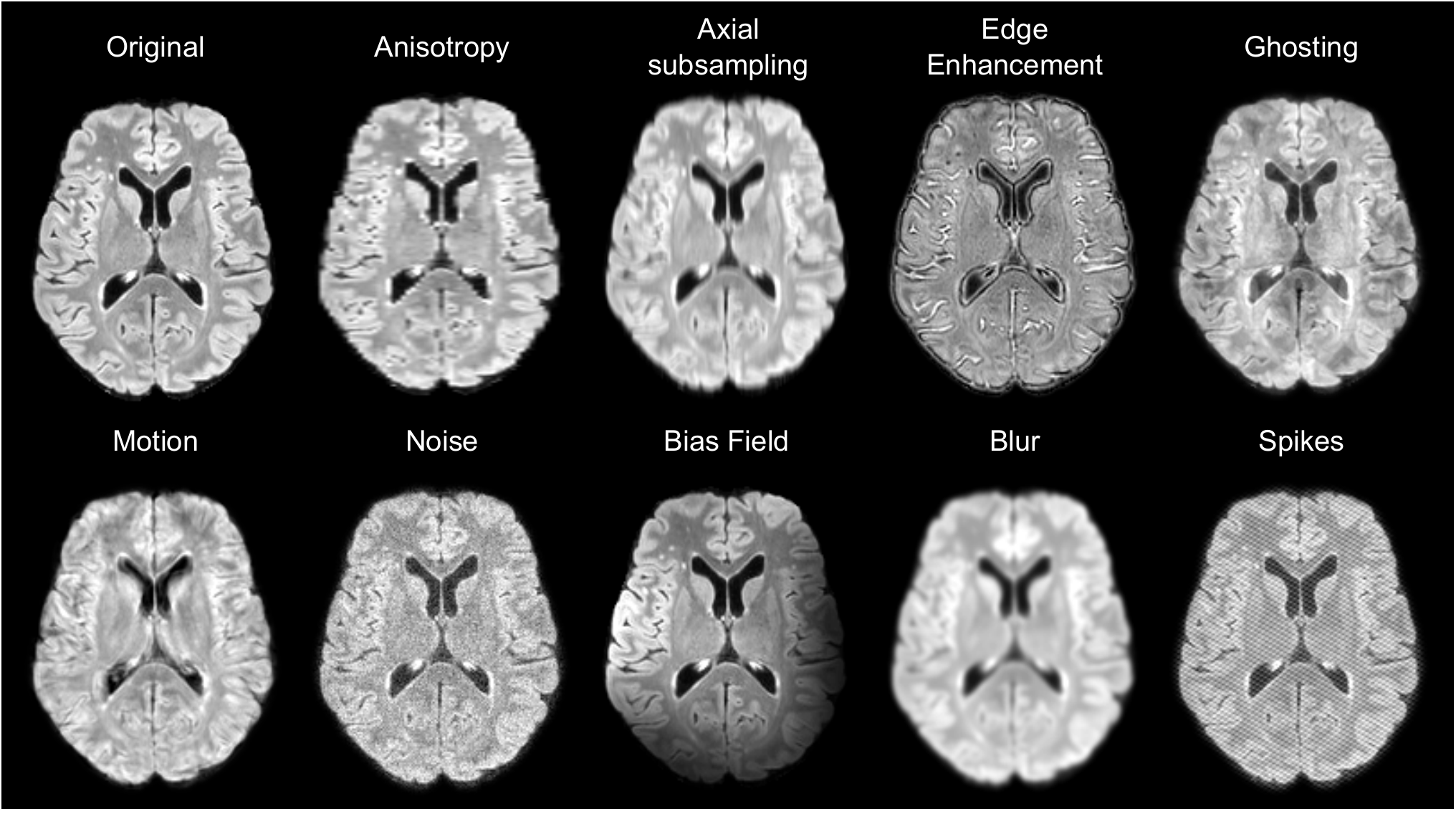}
      \caption{Examples of data augmentation applied on FLAIR images}
    \label{fig:augmentation}
\end{figure*}

\subsection{Data Augmentation} \label{IQDA}
The quality of the MRI greatly varies between datasets. The quality of the images depends on several factors such as signal-to-noise ratio, contrast-to-noise ratio, resolution, or slice thickness. 
To make our training stages robust to the diversity of image quality, Data Augmentation (DA) is used (see "DA" in Fig. \ref{fig:pipeline} and "Data Augmentation" in Fig. \ref{fig:synthesis}). 
We use an improved version of the data augmentation strategy proposed in \cite{kamraoui2022deeplesionbrain}, which simulates MRI quality disparity. 
During training, we simulate “on the fly” altered versions of 3D patches. We randomly introduce a set of alterations in the spatial and frequency space (k-space): Blur, edge enhancement, axial subsampling distortion, anisotropic downsampling, noise, bias-field variation, motion effect, MRI spike artifacts, and ghosting effect. Figure \ref{fig:augmentation} shows augmentation samples.

For the blur, a gaussian kernel is used with a randomly selected standard deviation (SD) ranging between $[0.5, 1.75]$. For edge enhancement, we use unsharp masking with the inverse of the blur filter. For axial subsampling distortion, we simulate acquisition artifacts that can result from the varying slice thickness. We use a uniform filter (a.k.a mean filter) along the axial direction with a size of $[1 \times 1 \times sz]$ where $sz \in {2,3,4}$. For anisotropic downsampling, the image is downsampled through an axis with a random factor ranging between $[1.5, 4]$ and upsampled back again with a B-spline interpolation.
For noise, we add to the image patch a Gaussian noise with $0$ mean and an SD ranging between $[0.02, 0.1]$. Bias-field variation is generated using the work of \cite{sudre2017longitudinal} that considers the bias field as a linear combination of polynomial basis functions. Motion effect has been generated based on the work of \cite{shaw2018mri}. The movements are simulated by combining in the k-space a sequence of affine transforms with random rotation and translation in the ranges $[-5, 5]$ degrees and $[-4, 4]mm$ respectively. Both MRI spike artifacts and ghosting effect have been generated with the implementation of \cite{perez2021torchio}.

\begin{table}[]
\centering
\caption{Summary of the used datasets. For each dataset, the object count (Obj. Count) and the total volume (Tot. Vol. $cm^3$) represent respectively the total number and the total volume in $cm^3$ of lesions or new lesions (depending on the task).}
\label{tab:dataset}
\resizebox{\textwidth}{!}{%
\begin{tabular}{c|ccccccc|}
\hline
Task                                                                               & Dataset                                                        &  Patients &  Time-point &  Raters & Obj. Count & Tot. Vol. ($cm^3$) & Clinical Site/ Scanners                                                                                                  \\ \hline
\multirow{3}{*}{\begin{tabular}[c]{@{}c@{}}MS Lesion \\ Segmentation\end{tabular}} & ISBI                                                           & 5          & 4-5          & 2   & 514 & 243     & Single-site                                                                                                              \\
                                                                                   & MSSEG' 16                                                      & 15         & 1            & 7     & 512 &  367 & Multi-site: 3 sites                                                                                                      \\
                                                                                   & In-house                                                       & 43         & 1            & 2 & 2391 & 1313        & Multi-site                                                                                                               \\ \hline
\multirow{2}{*}{\begin{tabular}[c]{@{}c@{}}New MS Lesion\\ Segmentation\end{tabular}} & \begin{tabular}[c]{@{}c@{}}MSSEG2 \\ Training-set\end{tabular} & 40         & 2            & 4   & 123 & 23     & \multirow{2}{*}{\begin{tabular}[c]{@{}c@{}}Multi-site:\\ 15 MRI scanners\\ ( GE scanners only in Test-set)\end{tabular}} \\
                                                                                   & \begin{tabular}[c]{@{}c@{}}MSSEG2 \\ Test-set\end{tabular}     & 60         & 2            & 4     & 174 &     60 &                                                                                                                          \\ \hline
\end{tabular}%
}
\end{table}

\subsection{Data}
Different datasets are used for the training and validation of the two tasks (see Table \ref{tab:dataset}).

\subsubsection{Single Time-point Datasets} \label{single_dataset}
For time-points synthesis (see \ref{synthesis}) and encoder pretraining (see \ref{encoder_pretraining}), we jointly used three datasets containing single time-ponints FLAIR and a lesion masks.
First, the ISBI \cite{carass2017longitudinal} training-set contains 21 FLAIR images with expert annotation done by two raters. Although the dataset is composed of longitudinal time-points from 5 patients, the provided expert annotations focus on the lesion mask of each time-point independently from the others and do not provide new lesion masks. Thus, we use the 21 images independently.    
Second, the MSSEG'16 training-set \cite{commowick2016msseg} contains 15 patients from 3 different clinical sites. Each FLAIR image is along with a consensus segmentation for MS lesions from seven human experts.
Third, our in-house \cite{coupe2018lesionbrain} dataset is composed of 43 subjects diagnosed with MS. The images were acquired with different scanners and multiple resolutions and their lesion masks have been obtained by two human experts.

All images were pre-processed using the lesionBrain pipeline from the volBrain platform \cite{manjon2016volbrain}. First, it includes image denoising \cite{manjon2010adaptive}. Second, an affine registration to MNI space is performed using the T1w modality, then the FLAIR is registered to the transformed T1w. Skull stripping and bias correction have been performed on the modalities, followed by the second denoising. Finally, the intensities have been normalized with kernel density estimation.

\subsubsection{Two Time-points Datasets}
The dataset provided by the MSSEG2-challenge \cite{challenge_dataset} is used to train our method.
The challenge dataset features a total of 100 MS patients. For each patient, two 3D FLAIR sequence time-points have been acquired spaced apart by a 1 to 3 years period.
The dataset has been split into 40 patients for training and 60 patients for testing.
A total of 15 different MRI scanners were used for the acquisition of the entire dataset. However, all images from GE scanners have been reserved only for the testing set to see the generalization capability of the algorithms.
Reference segmentation on these data was defined by a consensus of 4 expert neuroradiologists.

For preprocessing, the challenge organizers proposed a docker \footnote{https://github.com/Inria-Empenn/lesion-segmentation-challenge-miccai21/} built with the Anima scripts. It includes bias correction, denoising, and skull stripping. In addition, we added a registration step to the MNI space using a FLAIR template (\textit{i.e} the training and inference are performed in the MNI space, then the segmentation masks are transformed-back to the native space for evaluation).

Before challenge day, the testing set (the 60 patients) was not publicly available. Thus to test our methods (see Section \ref{ablation}), we defined an internal validation subset from the 40 challenge training data. From the 40 patients, 6 cases containing confirmed new lesions
were kept out from the training-set and were used as an internal test-set.
For the challenge evaluation (see Section \ref{challenge_eval}), the model submitted to the challenge organizers was trained on the entire MSSEG2 training-set.

\subsubsection{Dataset class imbalance}\label{dataset_imbalance}
 
Anomaly detection/segmentation tasks, such as MS lesion segmentation, suffer from class imbalance where the positive class is scarce (see \cite{johnson2019survey}). 
Herein, the MSSEG2-challenge \cite{challenge_dataset}  dataset is composed of 100 patients (40 for training and 60 for test) and all the MS Lesions Segmentation datasets combined account for 64 patients and 79 images.
Therefore, the number of image is similar. However, the class imbalance is highly different when evaluating the class imbalance using the number of objects to detect/segment (which represent MS lesions for the first task and new lesions for the second one) and their total volume for each dataset (see Table \ref{tab:dataset}). 
Indeed, we see that the MSSEG2-challenge datasets (especially training-set) suffer from more severe under-representation of the positive class. 
Consequently, it will be more difficult to train a model for New MS lesion segmentation than for the task of single time-point MS lesion segmentation. Furthermore, it shows that MS lesion segmentation datasets could significantly enrich the training of New MS lesion segmentation models. 

\subsection{Implementation Details}
First, all models are trained on 3D image patches of size $[64\times64\times64]$.
For the two time-points new lesion model, an ensemble of 5 networks (different training/validation data-split) is used. During inference, the consensus (prediction average) of the ensemble segmentation is used with a threshold of 0.5 to obtain the binary segmentation at the voxel level (new lesion voxel or not).

Second, the Dice-loss (soft DICE with probabilities as continuous values) is used as a loss function for the training of the single time-point MS lesion segmentation and the two time-points new lesion models. The mean-squared error is used as a loss function to train time-point synthesis models (inpainting and lesion generator models).

Finally, the experiments have been performed using PyTorch framework version 1.10.0 on Python version 3.7 of Linux environment with NVIDIA Titan Xp GPU 12 GB RAM. All models were optimized with Adam \cite{kingma2014adam} using a learning rate of 0.0001 and a momentum of 0.9.

\subsection{Validation Framework}
\subsubsection{Evaluation Metrics}
The assessment of a segmentation method is usually measured by a similarity metric between the predicted segmentation and the human expert ground truth. 

First, we use the Dice similarity coefficient defined as twice the intersection of the predicted and the ground truth positive regions over the sum of both regions. 
\begin{equation}
Dice=\frac{2 \times T P}{ (T P+F N) +(T P+F P) }\, \mathrm{,}
\end{equation}

where TP, FN, and FP represent respectively true positives, false negatives, and false positives.

Second, recent works (\textit{i.e.}, \cite{commowick2018objective}) question the relevance of classic metrics (Dice) compared to detection metrics, which are used for MS diagnostic and clinical evaluation of the patient evolution. Thus, in addition to the voxel-wise metrics, we also use lesion-wise metrics that focus on the lesion count. We use the lesion detection F1 ($LesF_{1}$) score defined as 
\begin{equation}
LesF_{1}=\frac{2 \times S_L \times P_L}{ (S_L+P_L) }\, \mathrm{,}
\end{equation}

where $S_L$ is lesion sensitivity, \textit{i.e.} the proportion of detected lesions and $P_L$ is lesion positive predictive value, i.e. the proportion of true positive lesions. 
For result harmonization with challenge organizers and participants, the same evaluation tool is used, \textit{i.e.} animaSegPerfAnalyzer \cite{commowick2018objective}.
All lesions that are smaller in size than $3mm^3$ are removed. For $S_L$, only ground-truth lesions that overlap at least $10\%$ with segmented volume are considered positive. For a predicted lesion to be considered positive for $P_L$ it has to be overlapped by at least $65\%$ and do not go outside by more than $70\%$ of the volume.

Finally, to jointly consider the different metrics (\textit{i.e.}, segmentation and detection performance), it would be convenient to aggregate them into a single score. Thus, we propose the average of DICE and $LesF_{1}$ (Avg. Score) as an aggregation score for comparing different metkhods. 

\subsubsection{Statistical Test}
To assert the advantage of a technique obtaining the highest average score, we conducted a Wilcoxon test (\textit{i.e.}, paired statistical test) over the lists of metric scores.
The significance of the test is established for a \textit{p}-value below 0.05. In the following tables, * indicates a significantly better average score when compared with the rest of the other approaches.

\section{Results}
Several experiments were conducted on our methods, including an ablation study and the comparison with state-of-the-art methods in competition during the challenge evaluation.

\subsection{Ablation Study}\label{ablation}
\begin{table}[]
\centering
\caption{The internal validation results for the ablation study. \cmark and \xmark symbolize using or not each contribution.  Bold values indicate the best result for a metric and * indicates that the advantage is statistically significant (Wilcoxon test) }
\label{tab:ablation}
\resizebox{300pt}{!}{%
\begin{tabular}{c|c|c|c|c|c|}
\hline
\begin{tabular}[c]{@{}l@{}} Transfer\\ Learning\end{tabular} & \begin{tabular}[c]{@{}l@{}}Time-point \\ Synthesis \end{tabular} & \begin{tabular}[c]{@{}l@{}} Data\\augm.\end{tabular} & \begin{tabular}[c]{@{}l@{}}Avg. \\Score\end{tabular} & DICE   & $LesF_{1}$  \\ \hline
\cmark & \cmark   & \cmark & \textbf{0.543*}     & \textbf{0.514*} & \textbf{0.573*}  \\
\cmark   & \xmark & \xmark & 0.483 & 0.480 & 0.486   \\
\xmark & \cmark & \xmark  &  0.501 &  0.461 & 0.541 \\
\xmark   & \xmark & \cmark &  0.477  & 0.464 & 0.488 \\         
\xmark & \xmark & \xmark  &    0.469 & 0.449 & 0.489  \\

\hline
\end{tabular}%
}
\end{table}
To evaluate each contribution of our training pipeline, Table \ref{tab:ablation} compares our full method with a baseline and other variations of our method on the internal validation dataset.
The baseline in this experiment was trained with real time-points only and by using a classic data augmentation composed of orthogonal rotations and mirroring.

First, when using only transfer learning on the top of the baseline, we measured an increase in DICE compared to the baseline but approximately the same $LesF_{1}$.
Second, when using only time-point synthesis pretraining on the top of the baseline, we obtained a significantly higher $LesF_{1}$ compared to the baseline and an increase in DICE. 
Third, when comparing the use of the proposed data augmentation, we see an increase in DICE but approximately the same $LesF_{1}$. 
Finally, when combining the transfer learning time-point synthesis pre-training, and the proposed data-augmentation, we obtained significantly higher scores in all metrics.

\subsection{Challenge Evaluation} \label{challenge_eval}
To evaluate our method on the challenge dataset, Table \ref{tab:my} compares it to the leader-board state-of-the-art methods. Results of the top performing methods were reported from challenge-day results.

Besides the top-performing methods, Table \ref{tab:my} also includes the expert raters performance to give an insight into the human performance. 
Their performance is measured compared to each other, contrary to the top methods that are evaluated using consensus segmentation. 
Raters $x$ Vs. $y$ means that we evaluate the performance of rater $x$ when considering rater $y$ segmentations as ground truth. Indeed, we consider that such a strategy can be more meaningful than the consensus segmentation in our case since the expert consensus already encodes the raters segmentation and thus is unfair when comparing to other strategies that did not participate in the consensus. 

First, from the Top 5 best-performing methods, LaBRI-IQDA \cite{kamraoui2021image} (our team's submission during the challenge-day) obtained the best score during the challenge. This method was similar to the proposed baseline with data augmentation.
Second, the proposed method (results obtained after challenge-day) obtained the highest $LesF_{1}$ and Average score. Moreover, these both scores are significantly better than all the listed state-of-the-art methods. The DICE score obtained by MedICL was not significantly better than the one obtained by our method.
Third, all but one (Empenn) leader-board automatic method obtained better DICE than raters segmentation. Our proposed method, LaBRI-IQDA, and MedICL even surpassed all raters in Average Scores.

\begin{table}[]
\centering
\caption{Results of MSSEG2-challenge \cite{challenge_dataset} evaluation. From top to bottom, the table shows the challenge raters agreement on the segmentation compared to each other, the leader-board results of the challenge-day top methods, and the result of the method described in this paper (obtained after challenge-day).
For automatic methods, bold values indicate the best result for a metric and * indicates that the advantage is statistically significant (Wilcoxon test)}
\label{tab:my}
\resizebox{350pt}{!}{%
\begin{tabular}{cc|c|c|c}
\hline
\multicolumn{2}{c|}{Experiment}                                          & Avg. Score      & DICE           & $LesF_{1}$      \\ \hline
\multicolumn{2}{c|}{Raters 1 Vs. 2}                                      & 0.466           & 0.426          & 0.507           \\
\multicolumn{2}{c|}{Raters 1 Vs. 3}                                      & 0.499           & 0.434          & 0.564           \\
\multicolumn{2}{c|}{Raters 1 Vs. 4}                                      & 0.434           & 0.382          & 0.486           \\ \hline 
\multirow{5}{*}{\rotatebox[origin=c]{90}{Challenge-day}} & LaBRI-IQDA \cite{kamraoui2021image}   & 0.507           & 0.498          & 0.515           \\
                               & MedICL \cite{zhang2021segmentation}   & 0.503           & \textbf{0.506} & 0.5             \\
                               & SNAC \cite{cabezas2021estimating}    & 0.496           & 0.484          & 0.513           \\
                               & Mediaire-B \cite{dalbis2021triplanar} & 0.489           & 0.436          & 0.541           \\
                               & Empenn \cite{masson2021nnunet}        & 0.478           & 0.423          & 0.532           \\[10pt] \hline 
\multicolumn{2}{c|}{The Proposed Method}                                 & \textbf{0.523*} & 0.495          & \textbf{0.550*} \\ \hline 
\end{tabular}%
}
\end{table}

\begin{figure*}[t]
    \centering
    \newcommand{\sz}{1\textwidth}
    \includegraphics[width=\sz]{"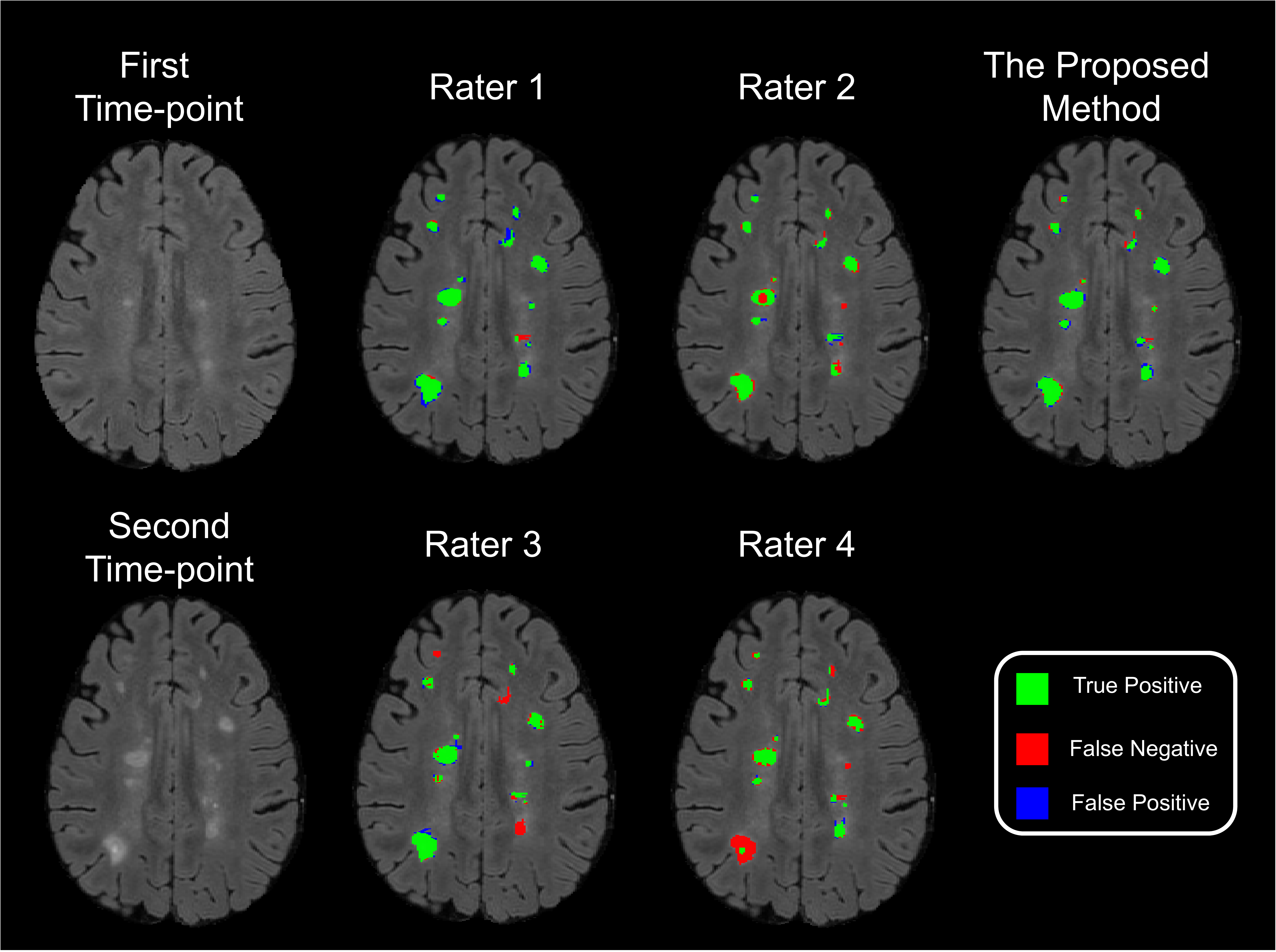"}
      \caption{The segmentation of the proposed method and the expert rater on a sample image from MICCAI 2021 - Longitudinal Multiple Sclerosis Lesion Segmentation Testing Dataset. The segmentations are compared against the consensus of the 4 raters using the colors: green, red, and blue to symbolize TP, FN, and FP regions of new lesions.}
    \label{fig:sample_segmentation}
\end{figure*}

Figure \ref{fig:sample_segmentation} shows the segmentation of new lesions of our proposed method. As a ground-truth reference, we compare the segmentation with the consensus segmentation of raters.
We also compare each rater segmentation against their consensus. 
From the five segmentation, we see that our segmentation is the most accurate with the consensus. 
Each of the human experts Rater 2, Rater 3, and Rater 4 missed one or multiple lesions when segmenting this sample.
While Rater 1 did not miss any lesions, we see that our segmentation is the closest to the consensus compared to his/her.

Overall, our method obtained the best result in the MSSEG2 challenge evaluation (during the challenge and after). Moreover, the result of the experiments showed that our segmentation is objective and can produce more accurate segmentations than human raters.   

\section{Discussion}

The transfer-learning from single time-point MS lesion segmentation task is an effective method to train the model for the task of two time-points new MS lesion segmentation even with a small dataset. Indeed, it enables to exploit the large available MS cross-sectional datasets compared to longitudinal datasets.
In our case, the encoder for the first task was compatible with the siamese-encoder of the second task and thus was used to extract MS-relevant features from the two time-points. Additionally, we used a learnable aggregation module for time-points feature combination. 
Besides, by freezing the encoder weights after the transfer-learning from the first to the second task, we ensure that the extracted features in the second task are dataset-independent from the second task dataset (smaller dataset). This independence ensures that the high performance of the proposed method is stable and generalizing.

Longitudinal time-points synthesis is an original approach on how to augment data diversity. It can be extended to other change detection tasks where longitudinal data are hard to acquire. According to the results of our experiments, this strategy turns out to be very effective when used as pretraining. Indeed, when the model is first pretrained with time-point synthesis, it is subject to a wider range of diversity, which aims to constrain the model to extract more generalizing features.

The proposed data augmentation method is an effective technique to make our learning process less dependent on MRI quality and acquisition artifacts. It simulates different acquisition conditions to enhance generalization.
Our data-augmentation comparison (see Table \ref{tab:ablation}) showed the proposed augmentation method contributes to segmentation accuracy in both internal validation and challenge evaluation (i.e. MRI from scanners not seen during training).

The ablation study performed using the internal validation process showed that each contribution, taken separately, enhanced the segmentation accuracy. It also showed that when combining all contributions, we achieved the best results. Similarly, the challenge evaluation showed that the proposed method achieved better results than the best-performing methods of the challenge.  


\section{Conclusion}
In this paper, we propose a training pipeline to deal with the lack of data for new MS lesion segmentation from two time-points. 
The pipeline encompasses transfer learning from single time-point MS lesion segmentation, pretraining with time-point synthesis, and data-augmentation adapted for MR images. 
Our ablation study showed that each of our contributions enhances the accuracy of the segmentation. 
Overall, our pipeline was very effective for new MS lesions segmentation (Best score in MSSEG2-challenge \cite{challenge_dataset}) and can be extended to other tasks that suffer from longitudinal data scarcity.

\section{Acknowledgements}
This work benefited from the support of the project DeepvolBrain of the French National Research Agency (ANR-18-CE45-0013). This study was achieved within the context of the Laboratory of Excellence TRAIL ANR-10-LABX-57 for the BigDataBrain project. Moreover, we thank the CNRS/INSERM for the DeepMultiBrain project. This study has also been supported by the PID2020-118608RB-I00 grants from the Spanish Ministerio de Economia, Industria Competitividad. The authors gratefully acknowledge the support of NVIDIA Corporation with their donation of the TITAN Xp GPU used in this research.
\newpage

\bibliographystyle{abbrv}
\bibliography{references}

\end{document}